\documentstyle[12pt,worldsci]{article}
 1
\def\be{\begin{eqnarray}}
\def\en{\end{eqnarray}}
\def\non{\nonumber}
\def\la{\langle}
\def\ra{\rangle}
\def\ep{\varepsilon}
\def\np #1 #2 #3 {Nucl.~Phys.~{\bf#1},\ #2 (#3)}
\def\pl #1 #2 #3 {Phys.~Lett.~{\bf#1},\ #2 (#3)}
\def\pr #1 #2 #3 {Phys.~Rep.~{\bf#1},\ #2 (#3)}
\def\prd #1 #2 #3 {Phys.~Rev.~{\bf#1},\ #2 (#3)}
\def\prl #1 #2 #3 {Phys.~Rev.~Lett.~{\bf#1},\ #2 (#3)}
\def\rmp #1 #2 #3 {Rev.~Mod.~Phys.~{\bf#1},\ #2 (#3)}
\def\zp #1 #2 #3 {Z.~Phys.~{\bf#1},\ #2 (#3)}
\def\ijmp #1 #2 #3 {Int.~J.~Mod.~Phys.~{\bf#1},\ #2 (#3)}
\begin{document}
%
%
\rightline{\vbox{\halign{&#\hfil\cr
&IP-ASTP-14-95\cr
&June 1995\cr
}}}
%
%
\begin{center}{{\bf Nonfactorizable Corrections to Hadronic Weak Decays of
Heavy Mesons\footnote{To appear in the proceedings of
{\it The International Symposium on Particle
Theory and Phenomenology,
Iowa State University, May 22-24, 1995.}}
\\}
\vglue 0.6cm
{Hai-Yang Cheng
\\}
\baselineskip=13pt
{\it Institute of Physics, Academia Sinica, Taipei, Taiwan 115, ROC
\\}
\vglue 0.1cm
ABSTRACT}
\end{center}
{\rightskip=3pc
 \leftskip=3pc
 \baselineskip=12pt
 \noindent
Status of nonfactorizable effects in exclusive hadronic weak decays of $D$ and
$B$ mesons is reviewed.
\vglue 0.6cm}
\section{Introduction}
     It is customary to make the factorization approximation to describe
the hadronic weak decays of mesons; that is, the meson decay amplitude is
dominated by the factorizable terms provided that final-state interactions
and nonspectator contributions are negligible.  The hadronic matrix
elements of the factorizable amplitude is factorized into the product of
two matrix elements of single currents, governed by decay constants and
form factors. Consider a generic two-body decay of a meson $M\to M_1+M_2$.
The factorizable parts of the decay amplitude can be classified into three
different categories \cite{bauer}:
\be
{\rm class~I~(external~W~emission)}: && a_1\la M_1|(\bar{q}_1q_2)|0\ra\la
M_2|(\bar{q}_3q_4)|M\ra,   \non \\
{\rm class~II~(internal~W~emission)}: && a_2\la M_2|(\bar{q}_3q_2)|0\ra\la
M_1|(\bar{q}_1q_4)|M\ra,
\en
and the third class involving decays in which $a_1$ and $a_2$ amplitudes
interfer. Meson $M_1$ in class I decays is generated from the charge current
$(\bar{q}_1q_2)\equiv\bar{q}_1\gamma_\mu(1-\gamma_5)q_2$, while $M_2$ in
the class II transition comes from the neutral current $(\bar{q}_3q_2)$. For
a given parent meson $M$, the parameters $a_1$ and $a_2$ are universal and
channel independent. For the QCD-corrected effective weak Hamiltonian
\be
{\cal H}_{\rm eff}\,\propto~c_1O_1+c_2O_2=c_1(\bar{q}_1q_2)(\bar{q}_3q_4)+
c_2(\bar{q}_1q_4)(\bar{q}_3q_2),
\en
the parameters $a_1,~a_2$ are related to the Wilson coefficient functions
$c_1$ and $c_2$ via
\be
a_1=c_1+c_2/N_c,~~~~a_2=c_2+c_1/N_c
\en
in the standard factorization approach, where the term proportional to
$1/N_c$ arises from the Fierz transformation.

     However, it is known that this factorization approach fails to describe
class II charmed decay modes, e.g., $D^0\to\bar{K}^0\pi^0,~D^+\to\phi\pi^+,
\cdots$, etc. \cite{cheng1}
For example, the ratio $\Gamma(D^0\to\bar{K}^0\pi^0)/\Gamma(D^0
\to K^-\pi^+)$ is predicted to be $\sim 0.02$, whereas experimentally it is
measured to be $0.51\pm 0.07$ \cite{PDG}. It was realized by several groups
\cite{fuku} that the discrepancy between theory and experiment is greatly
improved if Fierz transformed terms in (3) are dropped. It has been argued
that \cite{buras}
this empirical observation is justified in the so-called large-$N_c$
approach in which a rule of discarding subleading $1/N_c$ terms can be
formulated. There are several important implications of such a
approach: (i) The factorization hypothesis for nonleptonic meson decays is
justified in the limit of $N_c\to\infty$ since nonfactorizable contributions
are suppressed relative to the factorizable ones
by at least factors of $N_c$. (ii) Color suppression in the
class II transitions is no longer operative in charm decay as $a_2=c_2(m_c)
\approx -0.52$ and $a_1=c_1(m_c)\approx 1.26$. The sizable destructive
interference partially accounts for
the longer lifetime of $D^+$ relative to $D^0$. (iii) Contrary to the
meson case, the factorization approximation is not applicable to hadronic
baryon decays. The nonfactorizable $W$-exchange contributions, which manifest
as pole diagrams at the hadronic level, are no longer helicity and color
suppressed; color suppression is compensated by a combinatorial factor of
order $N_c$ stemming from the fact that the baryon contains $N_c$ quarks in
the large-$N_c$ limit.

    Though the new $1/N_c$ factorization improves substantially over the
standard factorization for charm decay, it cannot be a universal
approach for describing the nonleptonic weak decays of mesons. First, a
theory by itself should be able
to specify the regime where it is applicable. However, there is no kinematic
region where the $1/N_c$ expansion is guranteed to be valid.
Second, it fails to explain the constructive
interference recently observed in charged $B$ decays \cite{cleo}: $B^-\to D^0
\pi^-,~D^0\rho^-,~D^{*0}\pi^-$. Therefore,
whether or not the large-$N_c$ picture works is at best case by case
dependent. If it operates, there must exist some dynamical reason for the
suppression of the $1/N_c$ terms. This implies that nonfactorizable terms
should play an essential role and it is our purpose to examine such effects.

\section{Nonfactorizable effects in $D\to PP,~VP$ decays}
    Considering the class-I decay $D_s^+\to\phi\pi^+$ as an example and using
the Fierz identity
\be
O_{1,2}=\,{1\over N_c}O_{2,1}+\tilde{O}_{2,1},
\en
where $\tilde{O}_1={1\over 2}\bar{q}_1\gamma_\mu(1-\gamma_5)\lambda^a q_2
\bar{q}_3\gamma^\mu(1-\gamma_5)\lambda^aq_4$, we find
\footnote{Recall that in the standard picture the hadronic matrix element
of the operator $O$ is evaluated by considering the contributions of the
operator itself and its Fierz transformation.}
\be
\la\phi\pi^+|{\cal H}_W|D_s^+\ra &\propto& (c_1+{c_2\over N_c}+c_2\chi_1)\la
\pi^+|(\bar{u}d)|0\ra\la\phi|(\bar{s}c)|D_s^+\ra +c_1\la\phi\pi^+|\tilde{O}_2|
D_s^+\ra  \non  \\
&+& (c_1+{c_2\over N_c})\la\phi\pi^+|O_1|D_s^+\ra_{nf}+(c_2+{c_1\over N_c})
\la\phi\pi^+|O_2|D_s^+\ra_{nf},
\en
where $\chi_1=\la\phi\pi^+|\tilde{O}_1|D_s^+\ra/\la\phi\pi^+|O_1|D_s^+\ra_f$
and the subscript $nf$ denotes nonfactorizable corrections to the matrix
elements of $O_{1,2}$. The nonfactorizable terms $\la\phi\pi^+|O_{1,2}|D_s^+
\ra_{nf}$ and $\la\phi\pi^+|\tilde{O}_2|D_s^+\ra$ in (5) are usually ignored
in the literature.

To proceed, we will assume that the nonfactorizable contributions are
dominated by the color-octet current $\tilde{O}_1$. Consequently,
$\la\phi\pi^+|{\cal H}_W|D_s^+\ra\propto a_1^{\rm eff}\la\pi^+|(\bar{u}d)|0\ra
\la\phi|(\bar{s}c)|D_s^+\ra$ with
\be
a_1^{\rm eff}=c_1+c_2({1\over N_c}+\chi_1).
\en
Likewise, nonfactorizable terms for the class II decay modes amount to a
redefinition of $a_2$:
\be
a_2^{\rm eff}=c_2+c_1({1\over N_c}+\chi_2).
\en
(For convenience, we will drop the superscript ``eff" henceforth.)
The key point is that the amplitudes of $D,~B\to PP,~VP$ are governed by
a single form factor so that nonfactorizable contributions due to final-state
soft gluon effects can be lumped into the effective parameters $a_1$ and
$a_2$. Though we do not know how to perform first-principles calculations
of $\chi_{1,2}$, we do expect that\cite{cheng2}
\be
|\chi(B\to PP)|<|\chi(D\to PP) |{\ \lower-1.2pt
\vbox{\hbox{\rlap{$<$}\lower5pt\vbox{\hbox{$\sim$}}}}\ }
|\chi(D\to VP)|,
\en
as soft gluon effects become stronger when the relative momentum of the
final-state particles becomes smaller, allowing more time for significant
final-state interactions (FSI).

   Because of the presence of FSI and the nonspectator contributions, it is
generally not possible to extract the nonfactorization parameters $\chi_{1,2}$
except for a very few channels. Therefore, in order to determine
$a_1$ and especially $a_2$ we should focus on the exotic channels
and the decay modes with one single isospin
component where nonspectator
contributions are absent and FSI are presumably negligible.
{}From data we find that\cite{cheng3}
\be
\chi_2(D\to\bar{K}\pi) &\simeq & -0.36\,,~~~~
\chi_2(D\to\bar{K}^*\pi) \simeq  -0.61\,,   \non \\
\chi_2(D^+\to\phi\pi^+) &\simeq & -0.44\,,~~~~\chi_1(D_s^+\to\phi\pi^+) \simeq
-0.60\,,
\en
where we have assumed $\chi_1\simeq \chi_2$ for $D\to\bar{K}^{(*)}\pi$ decays.
Note that, as pointed out in Ref.[9], the
solutions for $\chi$ are not uniquely determined. For example, another
possible solution for $\chi_2(D\to\bar{K}\pi)$ is $-1.18\,$. To remove
the ambiguities, we
have assumed that nonfactorizable corrections are small compared to the
factorizable ones. We see from (9) that in general $\chi_{1,2}$ and hence
$a_{1,2}$ are not universal and they are channel dependent
and satisfy the relation $|\chi(D\to PP)|<|\chi(D\to VP)|$ as expected.
We also see that since $\chi_2(D\to \bar{K}\pi)$ is close to $-{1\over
3}$, it is evident that a large cancellation between $1/N_c$ and $\chi_2(D\to
\bar{K}\pi)$ occurs. This is the dynamic reason why the large-$N_c$ approach
operates well for $D\to\bar{K}\pi$ decay. However, this is no longer the
case for $D\to VP$ decays; the predicted decay rates for $D\to VP$
in the large-$N_c$ approach in general disagree with data\cite{cheng3}.
Therefore, we are led to conclude that the leading $1/N_c$ expansion
cannot be a universal approach for the nonleptonic weak decays of the meson.
However, the fact that substantial nonfactorizable effects which contribute
destructively with the subleading $1/N_c$ factorizable contributions are
required to accommodate the data of charm decay means that, as far as
charm decays are concerned, the large-$N_c$ approach greatly improves the
naive factorization method in which $\chi_{1,2}=0$; the former approach
amounts to having universal nonfactorizable terms $\chi_{1,2}=-1/N_c$.
\section{Nonfactorizable effects in $B\to PP,~VP$ decays}
   If the large-$N_c$ picture is a universal approach for hadronic weak
decays of mesons, one will expect that $a_2(B)\simeq c_2(m_B)\approx -0.26$.
However, CLEO data\cite{cleo} clearly indicate a constructive interference in
charged $B$ decays $B^-\to D^0\pi^-,~D^0\rho^-,~D^{*0}\pi^-$ and hence a
positive $a_2$. This is a very stunning observation since it has been widely
believed by most practationers in this field that the $1/N_c$ expansion
applies equally well to the weak decays of the $B$ meson.

  Using the heavy-flavor-symmetry approach for heavy-light form factors and
assuming a monopole extrapolation for $F_1,~A_0,~A_1$, a dipole behavior
for $A_2,~V$, and an approximately constant $F_0$, as suggested by QCD
sum-rule calculations and some theoretical arguments\cite{cheng4}, we found
from CLEO data that the variation of $a_{1,2}$ from $B\to D\pi$ to $D^*\pi$
and $D\rho$ decays is negligible and the combined average is\cite{cheng4}
\footnote{The number $a_2/a_1=0.23\pm 0.11$ given in the CLEO paper\cite{cleo}
is obtained by a global least squares fit of the modified Bauer-Stech-Wirbel
model\cite{neubert} to the CLEO data of $B\to D^{(*)}\pi(\rho)$. (An
individual fit of the same model to the data gives rise to the average
$a_2/a_1=0.33\pm 0.08\,$.) Our result $a_2/a_1=0.22\pm 0.06$ thus improves
the previous error analysis by a factor of 2.}
\be
a_1[B\to D^{(*)}\pi(\rho)] =\,1.01\pm 0.06\,,~~~
a_2[B\to D^{(*)}\pi(\rho)] =\,0.23\pm 0.06\,,
\en
where we have neglected FSI and nonspectator effects, an assumption which is
probably justified in $B$ decays. It follows that
\be
\chi_1[B\to D^{(*)}\pi(\rho)]\simeq 0.05\,,~~~~\chi_2[B\to D^{(*)}\pi(\rho)]
\simeq 0.11\,.
\en
Since $|c_2|<<|c_1|$, it is clear that the determination of $\chi_1$ is far
more uncertain than $\chi_2$. Evidently, soft gluon effects are less
significant in $B$ decays, as what expected [see (8)].
For $B\to\psi K$ decays, we found\cite{cheng4}
\be
\left|a_2(B\to\psi K)\right|=\,0.225\pm 0.016\,.
\en
We have argued that its sign is positive since $\chi_2(B\to\psi K)$ should not
deviate too much from $\chi_2(B\to D\pi)$. It has been advocated by
Soares\cite{soares} that an analysis of the long-distance contribution
of $B\to\psi K$ to the decay $B\to K\ell^+\ell^-$ can be used to remove the
sign ambiguity of $a_2$.

    Thus far the nonfactorizable effect is discussed at a purely
phenomenological level. It is very important to have a theoretical estimate of
such effects even approximately. So far all existing theoretical calculations
rely on the QCD sum rule. The first pioneering work is due to Blok and
Shifman\cite{blok1} who calculated soft gluon contributions and found that
$1/N_c$ factorizable terms and the soft gluon effect $\chi$ almost compensate
in all $D\to PP$
decays and in some decay modes of $D\to VP$. They have applied the same
approach to the class I decay $\bar{B}^0\to D^+\pi^-$ and obtained\cite{blok2}
$\chi_1(\bar{B}^0\to D^+\pi^-)\sim -0.5\,$. Working in the framework of the
light-cone QCD sum rule,
R\"uckl and his collaborators\cite{ruckl} found a large cancellation between
the Fierz $1/N_c$ term and the nonfactorizable contribution $\chi_2$. Most
recently, Halperin\cite{halp} extended the same calculation to the class II
decay $\bar{B}^0\to D^0\pi^0$ and found $\chi_2(\bar{B}^0\to D^0\pi^0)\sim
-0.35$ and hence a negative $a_2(B\to D\pi)$, which is in contradiction with
experiment. It appears that all present QCD sum-rule calculations tend
to imply that the rule of discarding $1/N_c$ terms seems to hold in class-I
and class-II decays of the $B$ meson. It is thus a great challenge to the
theorists to understand the origin of disagreement between theory and
experiment for the parameter $a_2(B\to D\pi)$. This tantalizing issue
should be clarified
and resolved in the near future. At present, lattice calulcations of
soft gluon effects are already available for $D\to \bar{K}\pi$ decay. An
extension of such a computation to class II decay modes of the $B$
meson is urged.
\section{Nonfactorizable effects in $B,~D\to VV$ decays}
   The study of nonfactorizable effects in $M\to VV$ decay is more
complicated as its general amplitude consists of three independent
Lorentz scalars:
\be
A[M(p)\to V_1(\ep_1,p_1)V_2(\ep_2,p_2)]\propto \ep^*_\mu(\lambda_1)
\ep^*_\nu(\lambda_2)(\hat{A}_1g^{\mu\nu}+\hat{A}_2p^\mu p^\nu+i
\hat{V}\epsilon^{\mu\nu\alpha\beta}p_{1\alpha}p_{2\beta}),
\en
where $\hat{A}_1,~\hat{A}_2,~\hat{V}$ are related to the form factors
$A_1,~A_2$ and $V$ respectively. Since {\it a priori} there is no reason to
expect that nonfactorizable terms weight in the same way to $S$-, $P$-
and $D$-waves, namely
$A_1^{nf}/A_1=A_2^{nf}/A_2=V^{nf}/V$, we thus cannot define
$\chi_1$ and $\chi_2$. Consequently, {\it it is in general not possible to
define an effective $a_1$ or $a_2$ for $M\to VV$ decays once nonfactorizable
effects are taken into account\cite{kamal}.}

  It was pointed out recently that there are two
experimental data, namely the production ratio $R\equiv \Gamma(B\to\psi K^*)/
\Gamma(B\to\psi K)$ and the fraction of longitudinal polarization $\Gamma_L/
\Gamma$ in $B\to\psi K^*$, which cannot be accounted for simultaneously by all
commonly used models within the framework of factorization\cite{gourdin}. The
experimental results are
\be
R=\,1.74\pm 0.39,~~~~{\Gamma_L\over\Gamma}=\,0.78\pm 0.07\,,
\en
where the latter is the combined average of the measurements by ARGUS,
CDF\cite{argus} and CLEO\cite{cleo}.
Irrespective of the production ratio $R$, all the existing models fail to
produce a large longitudinal polarization fraction\cite{gourdin}. This strongly
implies that the puzzle with $\Gamma_L/\Gamma$ can be resolved only by
appealing to nonfactorizable effects.
However, if the relation $A_1^{nf}/A_1=A_2^{nf}/A_2=V^{nf}/V$ holds, then an
effective $a_2$ can be defined for $B\to\psi K^*$ and the prediction of
$\Gamma_L/\Gamma$ will be the same as that in the factorization approach as
the polarization fraction is independent of $a_2$. As a result, {\it
nonfactorizable terms should contribute differently to $S$-, $P$- and
$D$-wave amplitudes if we wish to explain the observed $\Gamma_L/\Gamma$.}

{}From data of $B\to\psi K^*$ and $D\to\bar{K}^*\rho$ decays,
we found that the nonfactorizable terms in $M\to
VV$ decay in general satisfy the relation\cite{cheng3}
\be
{A_1^{nf}\over A_1}>{A_2^{nf}\over A_2}\,,~{V^{nf}\over V}\,.
\en
The difference between $B\to VV$ and $D\to VV$ decays stems from the fact
that $A_1^{nf}$ is positive in the former, while it is negative in charm
decay. Since the magnitude of $A_2^{nf}/A_2$ is larger than that of
$A_1^{nf}/A_1$ in $D\to\bar{K}^*\rho$ decay, it is evident that the assumption
of $S$-wave dominance for nonfactorizable terms fails in charm decay.
We thus urge experimentalists to measure the polarized decay rates
in the color- and Cabibbo-suppressed decay mode $D^+\to\phi\rho^+$ decay
to gain insight in the nonfactorizable effects in $D\to VV$ decay.
\vskip 1mm
\noindent{\bf Acknowledgment}~~This work was supported in part by
the National Science Council of ROC
under Contract No. NSC84-2112-M-001-014.
\bibliographystyle{unsrt}

\begin{thebibliography}{99}
%
\bibitem{bauer} M. Bauer, B. Stech, and M. Wirbel, \zp C34 103 1987 .
%
\bibitem{cheng1}
%
For a review, see e.g., H.Y. Cheng, \ijmp A4 495 1989 .
%
\bibitem{PDG}
Particle Data Group, \prd D50 1173 1994 .
%
\bibitem{fuku}
M. Fukugita, T. Inami, N. Sakai, and S. Yazaki, \pl 72B 237 1977 ; D. Tadi\'c
and J. Trampeti\'c, \pl 114B 179 1982 ; M. Bauer and B. Stech, \pl 152B 380
1985 .
%
\bibitem{buras}
A.J. Buras, J.-M. G\'erard, and R. R\"uckl, \np B268 16 1986 .
%
\bibitem{cleo}
CLEO Collaboration, M.S. Alam {\it et al.,} \prd D50 43 1994 .
%
\bibitem{cheng2}
H.Y. Cheng, \pl B335 428 1994 .
%
\bibitem{cheng3}
H.Y. Cheng, IP-ASTP-04-95 (revised) (1995).
%
\bibitem{soares}
J.M. Soares, \prd D51 3518 1995 .
%
\bibitem{cheng4}
H.Y. Cheng and B. Tseng, \prd D51 6259 1995 .
%
\bibitem{neubert}
M. Neubert, V. Riekert, Q.P. Xu, and B. Stech, in {\it Heavy Flavors},
edited by A.J. Buras and H. Lindner (World Scientific, Singapore, 1992).
%
\bibitem{blok1}
B. Blok and M. Shifman, Sov. J. Nucl. Phys. {\bf 45}, 35, 301,
522 (1987).
%
\bibitem{blok2}
B. Blok and M. Shifman, \np B389 534 1993 .
%
\bibitem{ruckl}
A. Khodjamirian and R. R\"uckl, MPI-PhT/94-26 (1994).
%
\bibitem{halp}
I. Halperin, \pl B349 548 1995 .
%
\bibitem{kamal}
A.N. Kamal and A.B. Santra, Alberta Thy-31-94 (1994).
%
\bibitem{gourdin}
M. Gourdin, A.N. Kamal, and X.Y. Pham, \prl 73 3355 1994 ;
R. Aleksan, A. Le Yaouanc, L. Oliver, O. P\`ene, and J.-C. Raynal,
\prd D51 6235 1995 .
%
\bibitem{argus}
ARGUS Collaboration, H. Albrecht {\it et al.,} \pl B340 217 1994 ;
CDF Collaboration, FERMILAB Conf-94/127-E (1994).
%
\end{thebibliography}

\end{document}